\documentclass[12pt]{article}

\usepackage{graphicx}

\usepackage{cite}

\advance \textheight by \topskip
\makeatletter
\@addtoreset{equation}{section}
 
\makeatother
\newcommand{\be}{\begin{equation}}
\newcommand{\ee}{\end{equation}}
\newcommand{\bea}{\begin{eqnarray}}
\newcommand{\eea}{\end{eqnarray}}
\newcommand{\dd}{\partial}

\def\>{\rangle}
\def\<{\langle}

\begin{document}

\title{
%\begin{flushright}
%{\small USACH-FM-01-02}\\[-0.4cm]
%{\small PM-01-07}\\[1cm]
%\end{flushright}
{\bf {}Field redefinition and renormalisability in scalar field theories}}

%Cubic root of translations in field theory }}

\author{
{\sf   N. Mohammedi} \thanks{e-mail:
nouri@lmpt.univ-tours.fr}$\,\,$${}$
%\thanks{This work was carried out at the Department of Applied Mathematics
%and Theoretical Physics, Cambridge, UK.} 
%{\sf  G. Moultaka }\thanks{e-mail:
%moultaka@lpm.univ-montp2.fr}$\,\,$${}^{b}$
% and
%{\sf M.~Rausch de Traubenberg}\thanks{e-mail:
%rausch@lpt1.u-strasbg.fr}$\,\,$$^{c}$\\
\\
%{\small {\it Universit\'e Fran\c{c}ois Rabelais de Tours,}}\\
{\small ${}${\it Laboratoire de Math\'ematiques et Physique Th\'eorique (CNRS - UMR 6083),}} \\
{\small {\it F\'ed\'eration Denis Poisson (FR CNRS 2964)}},\\
{\small {\it Universit\'e Fran\c{c}ois Rabelais de Tours,}}\\
{\small {\it Facult\'e des Sciences et Techniques,}}\\
{\small {\it Parc de Grandmont, F-37200 Tours, France.}}}
%\\ 
%{\small ${}^{b}${\it Laboratoire de Physique 
%Math\'ematique et Th\'eorique, CNRS UMR 5825, 
%Universit\'e Montpellier II,}}\\
%{\small {\it Place E. Bataillon, 34095 Montpellier,
%France}}\\
%{\small ${}^{c}${\it
%Laboratoire de Physique Th\'eorique, CNRS UMR  7085,
%Universit\'e Louis Pasteur}}\\
%{\small {\it  3 rue de
%l'Universit\'e, 67084 Strasbourg, France}}}
\date{}
\maketitle
\vskip-1.5cm

\vspace{2truecm}

\begin{abstract}

\noindent

We have addressed the issue of field redefinition in connection with 
renormalisability. Our study is restricted to theories of interacting scalar 
fields. We have, in particular, shown that if a theory is renormalisable
in the usual power-counting sense then it remains renormalisable in the same
sense after a change of variables. This is due to the use of the powerful method 
of the background field expansion. In the case of a single complex sclar field, 
it turns out that the determination of the counter-terms is much simpler when 
polar coordinates are used. We illustate this by carrying out a one-loop calculation 
in the latter case.

\end{abstract}

\newpage

%\section{The string effective action}
%\renewcommand{\theequation}{1.\arabic{equation}}   
%

\setcounter{equation}{0}

\section{Introduction}

The issue of coordinate transformations (also referred to as field redefinition
or reparametrisation invariance) in quantum field theory has been in the past
\cite{s1,s2,s3,s4,s5,s6,s7,s8}, and still is\cite{s9,s10,s11,s12,s13,s14,s15,s16,s17} 
a subject of renewed interest.  
The so-called ''equivalence theorem'' ensures, in principle \cite{ss1,ss2},  that the elements of 
the $S$-matrix
remain the same after a field redefinition. However, 
it is certainly not clear what becomes of the 
renormalisability of a theory when a coordinate transformation is carried out.
We should mention that the reparametrisation invariance of the effective action
has benn discussed in \cite{toms2}.
\par
{} For instance,
it is well-known that the Lagrangian density for a complex scalar field
as given by
\bea
{\cal L} &=& \dd_\mu \Phi^\star \dd^\mu \Phi 
-m^2\left(\Phi^\star\Phi\right)  
-\lambda\left(\Phi^\star\Phi\right)^2
\,\,\,\,\,\,.
\eea
is a renormalisable theory. In Cartesian coordinates where 
\be
\Phi={1\over\sqrt{2}} \left(\chi_1+i\chi_2\right)
\ee
this model 
describes two massive real fields $\chi_1$ and $\chi_2$ with a quartic interaction
as seen from the Lagrangian
\bea
{\cal L} &=& {1\over 2}\dd_\mu \chi_1 \dd^\mu \chi_1
+ {1\over 2}\dd_\mu \chi_2 \dd^\mu \chi_2
 -{m^2\over 2}\left(\chi_1^2+\chi_2^2\right) 
- {\lambda\over 4}\left(\chi_1^2+\chi_2^2\right)^2\,\,\,\,\,\,.
\label{cartesian}
\eea

\par
Let us now perform a field redefinition and 
parametrise the complex scalar field $\Phi$
as
\be
\Phi={1\over \sqrt{2}}\,\rho\,e^{i\theta/v}\,\,\,\,\,,
\ee
where $\rho$ and $\theta$ are the polar coordinates (real fields)
and $v$ is a dimensionful constant. The Lagrangian density
becomes
\bea
{\cal L} &=& {1\over 2}\dd_\mu \rho \dd^\mu \rho 
+{1\over 2}{\rho^2\over v^2}\,\dd_\mu \theta \dd^\mu \theta -{m^2\over 2}\,\rho^2 
-{\lambda\over 4}\,\rho^4  \,\,\,\,\,\,.
\label{polar-coord}
\eea
The first thing to notice in this parametrisation 
is that the kinetic term for the real field $\theta$ is not of the standard form.
Furthermore, it is not clear how one can see that the spectrum of the theory
contains two massive fields (as in Cartesian coordinates).
\par
If one insists on treating the Lagrangian (\ref{polar-coord}) as a conventional field theory,
then a change of variables is necessary. {}For instance, we could
change $\rho$ to
\be
\rho=v\,e^{\psi/v}\,\,\,\,\,,
\ee
where $\psi$ is the new field.
The Lagrangian (\ref{polar-coord}) becomes then
\bea
{\cal L} &=& e^{2\psi/v}\left[
{1\over 2}\dd_\mu \psi \dd^\mu \psi 
+{1\over 2}\dd_\mu \theta \dd^\mu \theta\right]
-{m^2v^2\over 2}\,e^{2\psi/v} - 
{\lambda v^4\over 4}\, e^{4\psi/v} \,\,\,\,\,\,.
\label{polar-coord-1}
\eea
Expanding the exponential, $e^{2\psi/v}= 1 + 2\psi/v + 2\psi^2/v^2 + \dots$, 
leads to a standard kinetic terms for the two real fields $\psi$ and 
$\theta$. In this parametrisation, the field $\theta$ is massless 
while all the mass is appropriated by the field $\psi$.

\par
The important feature of the Lagrangians (\ref{polar-coord}) and (\ref{polar-coord-1}) 
is that their
interaction parts are non-polynomial in nature. That is,  they involve
derivatives of the fields. Therefore, the usual power-counting argument 
of renormalisability does not apply here. 
 
\par
The  aim of this note is to address the issue of the renormalisability
of theories like the one in (\ref{polar-coord}) when a change of variables 
is carried out.
Our strategy is to treat the type of Lagrangians in (\ref{polar-coord}) and (\ref{polar-coord-1}) 
as a four-dimensional 
non-linear sigma model supplemented with a potential term and study 
their renormalisability. We use, for this purpose, the background field 
method which we will review in the next section.

\section{The covariant background field expansion}

The non-linear sigma model is defined as follows: Let ${\Sigma}$ denote 
the four-dimensional spacetime with coordinates $x^\mu$ and derivative $\dd_\mu$
and let ${\cal M}$ be a Riemannian manifold (the target space) with
metric $g_{ij}$. The field of the sigma model $\phi^i(x)$ is 
a map from ${\Sigma}$ to ${\cal M}$. The Lagrangian for 
the non-linear sigma model is
\bea
{L}\left(\phi\right)= {1\over 2}
g_{ij}\left(\phi\right)\dd_\mu \phi^i \dd^\mu \phi^j
\,\,\,\,.
\label{sigma}
\eea
The field $\phi^i(x)$ labels the coordinates of the target space ${\cal M}$.

\par

The background field expansion method \cite{bf1,bf2,bf3,bf4,bf5,bf6} consists in splitting
the field $\phi^i(x)$ of the non-linear sigma model as 
$\phi^i(x)=\varphi^i(x)+\pi^i(x)$, where $\varphi^i$ is the background field (a classical field)
and $\pi^i$ is the quantum fluctuation around this background field. 
The quantum field $\pi^i=\phi^i-\varphi^i$, being a difference between
two coordinates of the target space, does not lead to a covariant expansion 
of the non-linear sigma model. In order to respect the geometric nature of 
the non-linear sigma model we expand the action in terms of the 
quantun field $\xi^i$ instead. This field transforms as a vector on the target space and is 
defined as follows:
\par
Let $\sigma^i(x,s)$ be 
the unique geodesic joining  the two target space points $\varphi^i(x)$ and
$\varphi^i(x)+\pi^i(x)$. The affine parameter $s\in \left[0\,,\,1\right]$ 
parametrises this geodesic and we have the
interpolating conditions
\be
\sigma^i(x,s=0)=\varphi^i(x) \,\,\,\,\,\,\,\,\,\,\,{\rm and}\,\,\,\,\,\,\,\,\,\,\,
\sigma^i(x,s=1)=\phi^i(x)=\varphi^i(x)+\pi^i(x)
\,\,\,\,\,.
\ee
The geodesic equation is given by 
\be
{{\rm d}^2\sigma^i\over {\rm d}^2s}+\Gamma^i_{jk}\left(\sigma\right){{\rm d}\sigma^j\over {\rm d}s}
{{\rm d}\sigma^k\over {\rm d}s}=0\,\,\,\,,
\ee
where $\Gamma^i_{jk}\left(\sigma\right)$ are the Christoffel symbols corresponding to the target space metric
$g_{ij}\left(\sigma\right)$. Let $\xi^i_s(x,s)$ denote the tangent vector to the geodesic $\sigma^i(x,s)$.
In other words,
\be
\xi^i_s(x,s)={{\rm d}\sigma^i\over {\rm d}s}\,\,\,\,\,.
\ee
The quantum field $\xi^i$ that will enter in the covariant expansion is defined 
as the tangent vector to the geodesic at the point $\varphi^i(x)$. That is,
\be
\xi^i\left(x\right)=\left.\xi^i_s\left(x,s\right)\right|_{s=0}\,\,\,\,\,\,.
\ee
Since $\xi^i(x)$ transforms as a vector on the target space, the expansion of the 
action in terms of this field will be automatically covariant. 

\par
In order to obtain the covariant expansion, we start by extending the Lagrangian as
\be
L\left(s\right)={1\over 2}g_{ij}\left(\sigma\left(s\right)\right)\dd_\mu\sigma^i\left(s\right)
\dd^\mu\sigma^j\left(s\right)\,\,\,\,\,
\ee
so that $L\left(\phi\right)=L\left(s=1\right)$. We then expand $L(s)$ in powers 
of $s$ around $s=0$. We obtain
\bea
L\left(\varphi+\pi\right)=L\left(s=1\right)
&=&\left.\sum_{n=0}^\infty {1\over n!}{{\rm d}^nL\left(s\right)
\over 
{\rm d}s^n}\right|_{s=0}
%\nonumber \\
=\left.\sum_{n=0}^\infty {1\over n!}\left(\nabla_s\right)^nL\left(s\right)\right|_{s=0}\,\,\,\,\,,
\label{expansion}
\eea
where we have used the fact that $L\left(\sigma(x,s)\right)$ is a scalar to get the last equality.
Here $\nabla_s$ is the covariant derivative along the curve $\sigma^i(x,s)$.
Its acts on an arbitrary tensor $T^i_j$ as
\be
\nabla_s T^i_j={{\rm d}\over {\rm d}s}T^i_j +\Gamma^i_{kl}\left(\sigma\right)\xi_s^k T^l_j 
-\Gamma^l_{kj}\left(\sigma\right)\xi_s^k T^i_l \,\,\,\,.
\ee
Of course, the tensor $T^i_j$ could have more indices.
If $T^i_j$ is a tensor function of $\sigma^i(x,s)$ only then
\be
\nabla_s T^i_j\left(\sigma\right)=\xi^k_s\nabla_k T^i_j\left(\sigma\right)
=\xi^k_s\left[{\dd \over \dd \sigma^k}T^i_j+\Gamma^i_{kl}\left(\sigma\right) T^l_j 
-\Gamma^l_{kj}\left(\sigma\right)T^i_l\right]\,\,\,\,\,.
\label{tensor}
\ee
Here $\nabla_i$ is the usual covariant derivative with respect to $\Gamma^i_{jk}$.

\par
The expansion (\ref{expansion}) is evaluated using (\ref{tensor}) together with the formulae
\bea
\nabla_s\dd_\mu\sigma^i &=& \nabla_\mu{{\rm d}\sigma^i\over{\rm d}s}=\nabla_\mu \xi^i_s
\equiv \dd_\mu \xi^i_s +\Gamma^i_{jk}\left(\sigma\right)\dd_\mu \sigma^j\xi^k_s\,\,\,\,,
\nonumber \\
\nabla_s g_{ij}\left(\sigma\right) &=& 0\,\,\,\,,
\nonumber \\
\nabla_s {{\rm d}\sigma^i\over{\rm d}s} &=& \nabla_s \xi^i_s=0\,\,\,\,,
\nonumber \\
\nabla_s \nabla_\mu \xi^i_s &=& R^i_{jkl}\xi^j_s\xi^k_s\dd_\mu\sigma^l
\,\,\,\,.
\eea
Here $R^i_{jkl}$ is the Riemann tensor{\footnote {Our convention is 
$R^i_{jkl}=\dd_k\Gamma^i_{jl} +\Gamma^i_{km}\Gamma^m_{jl}-\left(k\leftrightarrow l\right)$.}}.
\par
The first few terms in the expansion of the Lagrangian $L\left(\phi\right)$ around
the background $\varphi^i$ are
\bea
L\left(\phi\right) &=&  
{1\over 2}g_{ij}\left(\varphi\right)\dd_\mu\varphi^i
\dd^\mu\varphi^j+g_{ij}\left(\varphi\right)\dd_\mu\varphi^i\nabla^\mu\xi^j 
+{1\over 2}g_{ij}\left(\varphi\right)\nabla_\mu\xi^i
\nabla^\mu\xi^j
\nonumber \\ 
&+& R_{iklj}\left(\varphi\right)\dd_\mu\varphi^i\dd^\mu\varphi^j
\xi^k\xi^l
+\dots\,\,\,\,\,\,\,\,\,,
\label{expansion-1}
\eea
where
\be
\nabla_\mu\xi^i=\dd_\mu\xi^i +\Gamma^i_{jk}\left(\varphi\right)\dd_\mu\varphi^j\xi^k\,\,\,\,.
\ee
and $R_{iklj}=g_{im}R^m_{klj}$.
\par
If our Lagrangian contains a potential term like
\be
L\left(\phi\right) =  
{1\over 2}g_{ij}\left(\phi\right)\dd_\mu\phi^i\dd^\mu\phi^j-V\left(\phi\right)\,\,\,\,\,
\ee
then the expansion of the potential is simply
\be
V\left(\phi\right)=V\left(\varphi\right)+\sum_{n=1}^{\infty}{1\over n!}
\nabla_{j_{1}}\dots\nabla_{j_{n}}V\left(\varphi\right)\xi^{j_{1}}\dots\xi^{j_{n}}\,\,\,\,\,\,.
\label{expan-V}
\ee

\par
Notice that the term $g_{ij}(\varphi)\dd_\mu\xi^i\dd^\mu\xi^j$
needed for the determination of the propagator for the quantum field 
$\xi^i$ is not of the standard form due the the presence of the non-constant
metric $g_{ij}(\varphi)$. The remedy to this is to define a new quantum field 
$\xi^a$ as 
\be
\xi^a=e^a_i\xi^i\,\,\,\,\,\, {\rm or} \,\,\,\,\,\,
\xi^i=E^i_a\xi^a\,\,\,\,\,,
\ee
where we have introduced the vielbiens $e^a_i$ such that
\be
g_{ij}=\eta_{ab}\,e^a_ie^b_j\,\,\,\,\,.
\ee
The constat matrix $\eta_{ab}$ is invertible and $E^a_i$ is the inverse
of $e^i_a$. That is, 
\bea
e^a_iE^i_b=\delta^a_b\,\,\,\,\,\,,\,\,\,\,\,\, E^i_a e^a_j=\delta^i_j
\,\,\,\,\,\,.
\eea
With this field redefinition the expansions (\ref{expansion-1}) becomes
\bea
L\left(\phi\right) &=&  
{1\over 2}g_{ij}\left(\varphi\right)\dd_\mu\varphi^i\dd^\mu\varphi^j
+\eta_{ab}\,e^a_i\dd_\mu\varphi^i{\cal D}^\mu\xi^b
+{1\over 2}\eta_{ab}\,{\cal D}_\mu\xi^a {\cal D}^\mu\xi^b
\nonumber \\ 
&+& R_{iabj}\left(\varphi\right)\dd_\mu\varphi^i\dd^\mu\varphi^j\xi^a\xi^b
 +\dots\,\,\,\,\,\,\,\,\,,
\eea
where
\bea
{\cal D}_\mu\xi^a &=& \dd_\mu\xi^a+\omega^a_{ib}\,\dd_\mu\varphi^i\xi^b\,\,\,\,,
\nonumber \\
\omega^a_{ib} &=& e^a_j\left(\dd_iE^j_b+\Gamma^j_{ik}E^k_b\right)
=e^a_j\nabla_iE^j_b\,\,\,\,\,
\eea
and $R_{iabj}=R_{iklj}E^k_aE^l_b$. The propagator is now computed 
from the term $\eta_{ab}\,\dd_\mu \xi^a\dd^\mu\xi^b$ which has the standard
form.
\par
Similarly, the expansion (\ref{expan-V}) of the potential is
\bea
V\left(\phi\right) &=& V\left(\varphi\right)
+\sum_{n=1}^{\infty}{1\over n!}V_{{a_{1}}\dots{a_n}}\left(\varphi\right)
\xi^{a_{1}}\dots\xi^{a_{n}}
\nonumber \\
V_{{a_1}\dots{a_n}}\left(\varphi\right) &\equiv& 
E^{j_{1}}_{a_{1}}\dots E^{j_{n}}_{a_{n}}
\nabla_{j_{1}}\dots\nabla_{j_{n}}V\left(\varphi\right)
\,\,\,\,\,\,.
\eea
It is clear that $V_{{a_1}\dots{a_n}}$ is symmetric under the exchange of any two 
indices. It is convenient to write this as
\bea
V_{{a_1}\dots{a_n}} =D_{a_{1}}D_{a_{2}}\dots D_{a_{n}}V\,\,\,\,,
\label{V_{...}}
\eea
where the derivative $D_a$ acts as
\be
D_aX_{bc}=E^i_a\left(\dd_iX_{bc}-\omega^d_{ib}X_{dc}-\omega^d_{ic}X_{bd}\right)\,\,\,
\ee
on an arbitrary tensor $X_{bc}$.

\section{Field redefinition and sigma model}

We start with a field theory as described by the Lagrangian
\bea
{\cal L} = {1\over 2}
\eta_{ab} \dd_\mu f^a \dd^\mu f^b 
-V\left(f\right)
\,\,\,\,,
\label{renorma}
\eea
where $\eta_{ab}$ is a constant metric and $f^a(x)$ is a set of fields. 
In the special case of the complex scalar 
theory in (\ref{cartesian}), we have  $f^a=\left(\chi_1\,,\,\chi_2\right)$
and $\eta_{ab}={\rm diag}\left(1\,,\,1\right)$.
In four dimensions, this theory is renormalisable in the usual 
power-counting sense (Dyson criterion) if the potential $V\left(f\right)$
is at most quartic in the fields $f^a$. 
\par
Let now $\phi^i$ denote another set of fields which parametrise the same theory
as the fields $f^a$. In other words, we have made a change of variables from
the field $f^a$ to the fields $\phi^i$. We may therefore write 
\be
f^a=f^a\left(\phi^i\right)\,\,\,\,\,\,\,,\,\,\,\,\,\,\,\
\phi^i=\phi^i\left(f^a\right)\,\,\,\,\,,
\ee
where we have assumed that the change of variables is invertible.
Under this field redefinition,  the Lagrangian (\ref{renorma}) becomes
\bea
{L}= {1\over 2}
g_{ij}\left(\phi\right)\dd_\mu \phi^i \dd^\mu \phi^j
-V\left(\phi\right)
\,\,\,\,,
\eea
where the metric $g_{ij}$ is given by
\bea
g_{ij}=\eta_{ab}\dd_if^a\dd_jf^b\,\,\,\,\,.
\eea
Here $\dd_a={\dd\over\dd f^a}$ and $\dd_i={\dd\over \dd\phi^i}$ and the
range of the indices $a,b,..$ is the same as the range of the indices
$i,j,\dots\,\,\,$.

\par
{}From the relations $\dd_i\phi^j=\delta^j_i$ and $\dd_a f^b=\delta^b_a$ 
together with the chain rule, we deduce that
\be
\dd_if^a\dd_a\phi^j=\delta^j_i\,\,\,\,\,\,,\,\,\,\,\,\,
\dd_a\phi^i\dd_i f^b=\delta^b_a\,\,\,\,.
\label{e-E}
\ee
Hence we could identify the vielbeins $e^a_i$ and their inverse $E^i_a$
with
\be
e^a_i=\dd_i f^a\,\,\,\,\,\,,\,\,\,\,\, E^i_a=\dd_a\phi^i\,\,\,\,.
\label{vielbeins}
\ee 
The inverse of the metric is $g^{ij}=\eta^{ab}\dd_a\phi^i\dd_a\phi^j$,
where $\eta^{ab}$ is the inverse of $\eta_{ab}$. 
The Christoffel connection is then given by
\be
\Gamma^i_{jk}={1\over 2}g^{il}\left(\dd_jg_{lk}+ \dd_kg_{lj}-\dd_lg_{jk}\right)
=\dd_j\dd_kf^a\dd_a\phi^i\,\,\,\,\,
\ee
and all the components of the Riemann tensor $R^i_{jkl}$ vanish.
\par
Using the chain rule and the relations in (\ref{e-E}), 
we find that the spin connection, 
$\omega^a_{ib} = e^a_j\left(\dd_iE^j_b+\Gamma^j_{ik}E^k_b\right)$, vanishes. Indeed,
\bea
\omega^a_{ib} &=& \dd_i\dd_jf^a\dd_b\phi^j+ 
\dd_jf^a\dd_if^c\dd_c\dd_b\phi^j
\nonumber \\
&=& \dd_i\left(\dd_jf^a\dd_b\phi^j\right) - \dd_jf^a\dd_c\dd_b\phi^j\dd_if^c + 
\dd_jf^a\dd_if^c\dd_c\dd_b\phi^j = 0
\,\,\,\,\,.
\label{omega}
\eea
We also need the expression of the tensor $V_{a_{1}\dots a_{n}}$ as defined in (\ref{V_{...}}). This is
found to be 
\bea
V_{ab\dots c}=\dd_a\dd_b\dots\dd_c\,V\,\,\,\,\,.
\eea
If we assume that {\underline{ $V$ is at most quartic in the fieds}}, then 
the background field expansion, \underline{{\it to all order}} in the quantum field $\xi^a$,  
of the Lagrangian yields
\bea
L &=&
{1\over 2}
g_{ij}\left(\varphi\right)\dd_\mu \varphi^i \dd^\mu \varphi^j
-V\left(\varphi\right)
\nonumber \\
&+& \eta_{ab}\,e^a_i\dd_\mu\varphi^i\dd^\mu\xi^b
+{1\over 2}\eta_{ab}\,\dd_\mu\xi^a \dd^\mu\xi^b
\nonumber \\
&-&\dd_aV\xi^a -{1\over 2}\dd_a\dd_bV\xi^a\xi^b -{1\over 6}\dd_a\dd_b\dd_c V\xi^a\xi^b\xi^c
-{1\over 24}\dd_a\dd_b\dd_c\dd_d V\xi^a\xi^b\xi^c\xi^d
\,\,\,\,\,.
\label{all-orders}
\eea
We notice that the resulting expansion is at most quartic in the quantum 
field $\xi^a$. It is then clear that the theory is renormalisable in 
the usual power-counting sense. We will illustrate this by an example
in the following section.

\section{An example: the interacting complex scalar field in polar coordinates}

In the notation of the previous sections, we have
$f^a=\left(\chi_1\,,\,\chi_2\right)$ and the constant 
metric $\eta_{ab}={\rm diag}\left(1\,,\,1\right)$.
The new set of fields are $\phi^i=\left(\rho\,,\,\theta\right)$
and we have made the change of variables
\bea
\chi_1=\rho\cos\left(\theta/v\right)\,\,\,\,\,,\,\,\,\,\,
\chi_2=\rho\sin\left(\theta/v\right)\,\,\,\,.
\label{chi}
\eea
The Lagrangian (\ref{polar-coord}) is a non-linear sigma model with a 
metric $g_{ij}$ given by 
\be
g_{ij}= \left(
\begin{array}{cc}
1 & 0   \\
0 & {\rho^2/v^2} \end{array}
\right)\,\,\,\,\,.
\ee
The non-vanishing components
of its Christoffel symbols are
\bea
\Gamma^1_{22}=-{\rho\over v^2}\,\,\,\,\,\,,\,\,\,\,\,
\Gamma^2_{12}={1\over \rho}\,\,\,\,\,.
\eea

\par
Using (\ref{vielbeins}) and (\ref{chi}), the vielbeins $e^a_i$ and their inverses $E^i_a$ are{\footnote{
As matrices, $e^a_i$ and $E^i_a$ should be read as $e_{ia}$ and $E_{ai}$, respectively.}}
\be
e^a_i= \left(
\begin{array}{cc}
\cos(\theta/v) & \sin(\theta/v)   \\
-{\rho\over v}\,\sin(\theta/v) & {\rho\over v}\,\cos(\theta/v) \end{array}
\right)
\,\,\,\,,\,\,\,\,
E^i_a= \left(
\begin{array}{cc}
\cos(\theta/v) &-{v\over \rho}\, \sin(\theta/v)   \\
\sin(\theta/v) & {v\over \rho}\,\cos(\theta/v) \end{array}
\right)
\,\,\,\,\,.
\ee
Acoording to (\ref{omega}), all the components of the spin connection 
$\omega^a_{ib}$ vanish.

\par
The background field expansion, {\it to all order} in the quantum field $\xi^a$,  
of the Lagrangian (\ref{polar-coord}) yields
\bea
L\left(\varphi\,,\,\xi\right) &=&
{1\over 2}\dd_\mu \rho \dd^\mu \rho 
+{1\over 2}{\rho^2\over v^2}\,\dd_\mu \theta \dd^\mu \theta 
-\left({m^2\over 2}\,\rho^2 + 
{\lambda\over 4}\,\rho^4 \right) 
\nonumber \\
&+& \eta_{ab}\,e^a_i\dd_\mu\varphi^i\dd^\mu\xi^b
+{1\over 2}\eta_{ab}\,\dd_\mu\xi^a \dd^\mu\xi^b
\nonumber \\
&-&V_a\xi^a -{1\over 2}V_{ab}\xi^a\xi^b -{1\over 6}V_{abc}\xi^a\xi^b\xi^c
-{1\over 24}V_{abcd}\xi^a\xi^b\xi^c\xi^d
\,\,\,\,\,.
\eea
Here we have used $\varphi^i=\left(\rho\,,\,\theta\right)$ to denote also
the background fields. 

\par
The non-vanishing components of the symmetric tensor $V_{{a_1}\dots{a_n}}$ are
\bea
V_1 &=&\rho\left(m^2+\lambda\rho^2\right)\cos(\theta/v)\,\,\,\,\,\,\,\,,\,\,\,\,\,\,
\nonumber \\ 
V_2 &=& \rho\left(m^2+\lambda\rho^2\right)\sin(\theta/v)\,\,\,\,\,\,\,\,,
\nonumber \\
V_{11} &=& m^2 + \lambda\rho^2\left(1+2\cos(\theta/v)^2\right)
\,\,\,\,\,\,\,\,,\,\,\,\,\,\,
\nonumber \\
V_{12} &=& 2\lambda\rho^2\cos(\theta/v)\sin(\theta/v)
\,\,\,\,\,, 
\nonumber \\
V_{22}  &=& m^2 + \lambda\rho^2\left(1+2\sin(\theta/v)^2\right)
\,\,\,\,\,\,\,\,,\,\,\,\,\,\,
\nonumber \\
V_{111} &=& 6\lambda \rho\cos(\theta/v)\,\,\,\,\,\,\,\,\,\,\,,\,\,\,\,\,\,\,\,\,
\nonumber \\
V_{112} &=& 2\lambda \rho\sin(\theta/v)\,\,\,\,,
\nonumber \\
V_{122} &=& 2\lambda \rho\cos(\theta/v)\,\,\,\,,
\nonumber \\
V_{222} &=& 6\lambda \rho\sin(\theta/v)\,\,\,\,\,\,\,\,\,\,\,,\,\,\,\,\,\,\,\,\,
\nonumber \\
V_{1111} &=& 6\lambda \,\,\,\,\,\,\,\,\,\,\,,\,\,\,\,\,\,\,\,\,
\nonumber \\
V_{1122} &=& 2\lambda 
\,\,\,\,\,\,\,\,\,\,\,,\,\,\,\,\,\,\,\,\,
\nonumber \\
V_{2222} &=& 6\lambda \,\,\,\,\,. 
\eea
Let us now see what happens at the one-loop level in perturbation theory.

\noindent
\vskip 1.0cm
\noindent
{\bf\Large{One-loop renormalisation}:}
\vskip 1.0cm

\par
The generating functional for connected Green's functions,
$W\left[J\right]$,  is defined in the usual way by
\bea
Z\left[J\right]=e^{iW\left[J\right]}=
N\int\left[{\rm d}\xi\right]e^{i\left[S\left(\varphi\,,\xi\right)
+J_a\xi^a\right]}\,\,\,\,\,,
\eea
where $N$ is normalising factor and 
\be
\left[{\rm d}\xi\right]=\prod_x\prod_{a=1}^{D}
\sqrt{g\left(x\right)}\,{\rm d}\xi^a\,\,\,\,
\ee
is the coordinate independent measure.

\par
The Feynman propagator is computed from the free part of the action
\be
L_{\rm {free}}\left(\xi\right)={1\over 2}\eta_{ab}\dd_\mu\xi^a\dd^\mu\xi^b 
- {1\over 2}m^2\eta_{ab}\xi^a\xi^b\,\,\,\,\,.
\ee
The mass term comes from $V_{11}$ and $V_{22}$. This propagator
is given by
\bea
\Delta^{ab}_{{\rm F}}\left(x-y\right)=
{\eta^{ab}}\,{1\over(2\pi)^4}\int{e^{-ik.(x-y)}\over k^2-m^2}\, {\rm d}^4k\,\,\,
\eea
and satisfies
\be
\eta_{ac}\left(\dd_\mu\dd^u+m^2\right)\Delta^{cb}_{\rm {F}}
\left(x\right)=-\delta^a_b\delta^4\left(x\right)
\,\,\,\,\,.
\ee

\par
The loop expansion in terms of Feynman graphs is generated using 
the Dyson-Wick perturbation theory. This is obtained from
\bea
e^{iW\left(\varphi\right)}=<0|\exp\left[i\int{\rm d}^4x\,L_{{\rm int}}
\left(\varphi\,,\,\xi\right)\right]|0>
\,\,\,\,\,,
\eea
where
\be
L_{{\rm int}}
\left(\varphi\,,\,\xi\right)=L
\left(\varphi\,,\,\xi\right)-L_{{\rm free}}
\left(\xi\right)\,\,\,\,.
\ee
We then expand $\exp\left[i\int{\rm d}^4x\,L_{{\rm int}}
\left(\varphi\,,\,\xi\right)\right]$ 
and calculate the vacuum graphs by considering all the
possible Wick contractions involving the quantum field $\xi^a(x)$.
The background field $\varphi$ is treated as an external field.  

\par
We use dimensional regularisation to isolate the divergences 
in Feynman integrals. The dimension of spacetime is assumed
to be $d=4-\epsilon$. The original Lagrangian (\ref{polar-coord})
is  extended to $d$ dimensions as
\bea
L =
{1\over 2}\dd_\mu \rho \dd^\mu \rho 
+{1\over 2}{\rho^2\over v^2}\,\dd_\mu \theta \dd^\mu \theta 
-\left({m^2\over 2}\,\rho^2 + 
{\mu^{4-d}\lambda\over 4}\,\rho^4 \right)\,\,\,\,\,, 
\label{L-mu}
\eea
where $\mu$ is an arbitrary mass parameter. In $d$ dimensions
$\rho$ has a mass dimension equal to ${1\over 2}d-1$ while 
$\theta$ has a mass dimension equal to $1$. The coupling constant
$\lambda$ is kept dimensionless in $d$ dimensions. 

\par
The first divergent diagram  is shown in figure 1 and contributes
\begin{figure}[thp]
\begin{center}
\includegraphics[scale=0.8]{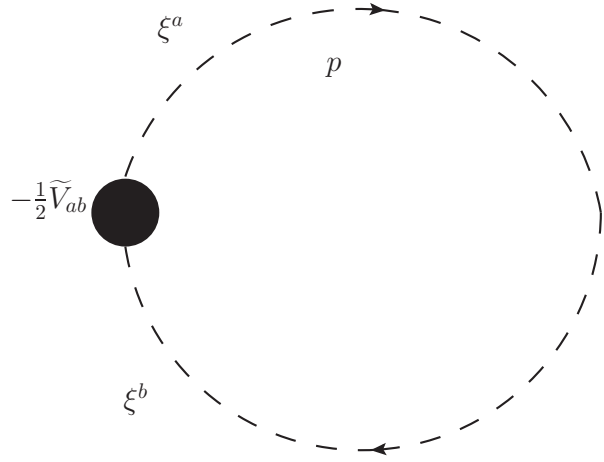}
\end{center}
\caption[]{One loop divergent diagram with one vertex ${\widetilde V}_{ab}$.}
\label{graph1}
\end{figure}

\bea
I_1=\left(i\right)^2\times\left(-{1\over 2}\right)
{\widetilde{V}}_{ab}\eta^{ab}\mu^{4-d}\int{{\rm d}^dp\over(2\pi)^d}\,{1\over p^2-m^2}\,\,\,\,,
\eea
where ${\widetilde{V}}_{11}=V_{11}-m^2$, ${\widetilde{V}}_{22}=V_{22}-m^2$
and ${\widetilde{V}}_{12}=V_{12}$.
The divergent part of the integral is (see for example \cite{ryder})
\be
I_1=-2\lambda\rho^2\left({im^2\over 8\pi^2\epsilon}+\,\,\,{\rm finite}\right)\,\,\,\,.
\ee

\par
The second divergent graph 
is drawn in figure 2 and gives
\begin{figure}[thp]
\begin{center}
\includegraphics[scale=0.8]{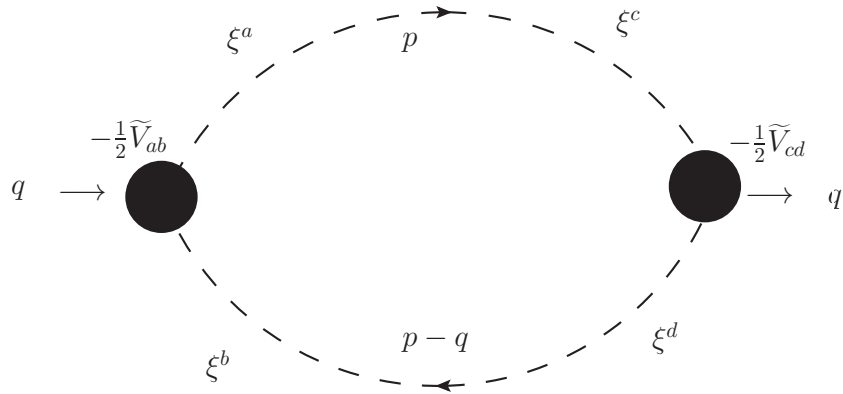}
\end{center}
\caption[]{One loop divergent diagram with two vertices ${\widetilde V}_{ab}$ and ${\widetilde V}_{cd}$.}
\label{graph2}
\end{figure}

\bea
I_2 &=& {\left(i\right)^4\over 2}\times
\left(-{1\over 2}\right)^2
{\widetilde V}_{ab}{\widetilde V}_{cd}\left(\eta^{ac}\eta^{bd}+\eta^{ad}\eta^{bc}\right)
\nonumber \\
 &\times& 
\left(\mu^2\right)^{4-d}
\int{{\rm d}^dp\over(2\pi)^d}\,{1\over \left(p^2-m^2\right)}{1\over \left[\left(p-q\right)^2-m^2\right]}\,\,\,\,.
\eea
The infinite part of the integral is extracted (see for example \cite{ryder}) and we find
\bea
I_2={5\over 2}\lambda^2\rho^4\left({i\mu^\epsilon\over 8\pi^2\epsilon} +\,\,\,{\rm finite}\right)\,\,\,\,.
\eea

\par
Green's functions are then rendered finite by adding to the
Lagrangian (\ref{L-mu}) the counter-term Lagrangian
\bea 
L_{{\rm CT}} =
{A\over 2}\dd_\mu \rho \dd^\mu \rho 
+{B\over 2}{\rho^2\over v^2}\,\dd_\mu \theta \dd^\mu \theta 
-\left({C\over 2 }\,\rho^2 + 
{D\mu^{\epsilon}\lambda\over 4}\,\rho^4 \right)\,\,\,\,\,.
\eea
This leads to the bare Lagrangian
\bea
L_{\rm B} =L+L_{{\rm CT}}=
{1\over 2}\dd_\mu \rho_{\rm B} \dd^\mu \rho_{\rm B}
+{1\over 2}{\rho^2_{\rm B}\over v^2_{\rm B}}\,\dd_\mu \theta_{\rm B} \dd^\mu \theta_{\rm B} 
-\left({m^2_{\rm B}\over 2}\,\rho^2_{\rm B} + 
{\lambda_{\rm B}\over 4}\,\rho_{\rm B}^4 \right)\,\,\,\,\,.
\eea
The bare quantities are defined as
\bea
\rho_{\rm B} &=&\sqrt{Z_{\rho}}\,\rho \,\,\,\,\,\,,\,\,\,\,\,\,
Z_\rho=1+A\,\,\,\,\,\,,
\nonumber \\
m_{\rm B} &=&{Z_m}m \,\,\,\,\,\,,\,\,\,\,\,\,
Z_{m}^2={m^2+C\over m^2(1+A)}\,\,\,\,\,\,,
\nonumber \\
\lambda_{\rm B} &=&\mu^\epsilon{Z_\lambda}\lambda \,\,\,\,\,\,,\,\,\,\,\,\,
Z_\lambda={1+D\over (1+A)^2}\,\,\,\,\,\,,
\nonumber \\
\theta_{\rm B} &=&\sqrt{Z_{\theta}}\,\theta
\,\,\,\,\,\,,\,\,\,\,\,\, v_{\rm B} ={Z_v}v
\,\,\,\,\,\,,\,\,\,\,\,\,{Z_\rho Z_\theta\over Z_v^2}=1+B\,\,\,\,.
\eea

\par
At the one loop level, we have
\be
C=-{\lambda m^2\over 2\pi^2\epsilon}\,\,\,\,\,\,\,\,,\,\,\,\,\,\,\,\,
D={5\lambda \over 4\pi^2\epsilon}
\,\,\,\,\,\,\,\,,\,\,\,\,\,\,\,\,
A=B=0
\,\,\,\,\,\,.
\ee
This leads to $Z_\rho=Z_\theta=Z_v=1$. The one-loop result is precisely 
the one found in the literature (see for example \cite{srednicki}).

\par
In conclusion, we have shown that a power-counting renormalisable scalar field theory
maintains this property in another reparametrisation of the fields.
The question of field redefinitions is much more important in theories involving gauge
fields. This is due to the fact that in some cases (like the Higgs model) 
one could work with gauge invariant 
variables which would eliminate the Faddeev-Popov ghosts. It is then crucial 
to see what happens to the theory under such change of variables.
Some progress has already been made in this direction \cite{toms1,sonoda1,sonoda2,masson}.

\smallskip
\smallskip
\smallskip
\bigskip

\noindent $\underline{\hbox{\bf Acknowledgments}}$: I am very greatful to 
David J. Toms for answering some questions regarding this work.
I would like also to thank Ian Jack and John Gracey for correspondence.

\end{document}